# Electrodynamics – molecular dynamics simulations of the stability of Cu nanotips under high electric field

**Mihkel Veske[1,2], Stefan Parviainen[1], Vahur Zadin[1,2], Alvo Aabloo[2] and Flyura Djurabekova[1]**

[1]Department of Physics and Helsinki Institute of Physics, University of Helsinki, PO Box 43 (Pietari Kalmin katu 2), 00014 Helsinki, Finland

[2]Intelligent Materials and Systems Lab, Institute of Technology, University of Tartu, Nooruse 1, 50411 Tartu, Estonia

E-mail: mihkel.veske@helsinki.fi

**Abstract**

The shape memory effect and pseudoelasticity in Cu nanowires is one possible pair of mechanisms that prevents high aspect ratio nanosized field electron emitters to be stable at room temperature and permits their growth under high electric field. By utilizing hybrid electrodynamics – molecular dynamics simulations we show that a global electric field of 1 GV/m or more significantly increases the stability and critical temperature of spontaneous reorientation of nanosized <100> Cu field emitters. We also show that in the studied tips the stabilizing effect of an external applied electric field is an order of magnitude greater than the destabilization caused by the field emission current. We detect the critical temperature of spontaneous reorientation using the tool that spots the changes in crystal structure. The method is compatible with techniques that consider the change in potential energy, has a wider range of applicability and allows pinpointing different stages in the reorientation processes.

## 1. Introduction

With the aim to challenge the frontiers of high energy physics, the European Organization for Nuclear Research (CERN) has initiated a new Compact electron-positron Linear Collider (CLIC) project [1]. The collision energies of leptons in CLIC are planned in the range of 0.5-5 TeV with optimal performance at 3 TeV [2]. To achieve such energies cost-efficiently, the particles must be accelerated by electric fields up to 300 MV/m [3] near the surface of the accelerating structures made out of copper. Using such high fields leads to vacuum breakdowns in the system. To ensure the proper behavior of the accelerator, the probability of those breakdowns must be held below $3 \cdot 10^{-7}$ pulse$^{-1}$m$^{-1}$ [4].

The triggering mechanism of vacuum breakdowns is not entirely clear yet, but experiments have shown a significant field enhancement near copper surfaces, that could lead to breakdown events via positive feedback effect [5]. Such an enhancement could be caused, for instance, by Cu nanotips with aspect ratio of 30-140 [5] whose formation mechanisms have been studied in [6]–[10]. However, such tips have not yet been directly observed on Cu surfaces [5], although there is a report about the formation of field emitters with aspect ratio 3.5-15 on Nb surfaces [11].

The lack of experimental observations of Cu nanotips could be explained by their natural tendency to collapse when no external force is acting on them. In long time scales the collapse is driven mostly by diffusion [12], [13] but in face-centered cubic (FCC) and body-centered cubic (BCC) nanowires the shape memory effect was shown to cause drastic reductions of major dimension within very short times [14]–[18]. In [14], the authors showed by means of molecular dynamics (MD) simulations that a Cu nanowire with <100> crystallographic orientation in longitudinal and {001} in lateral direction (denoted hereafter as <100>/{001}) reoriented into a <110>/{111} configuration in a matter of picoseconds if no external force acted on them. Such a phenomenon is strongly temperature-dependent due to shape memory effect (SME) – there exists a critical temperature $T_{cr}$ below which the wire was seen to be stable within the time interval simulated by MD and above it the reorientation occurred within a few picoseconds. The reorientation of <100> Cu nanowire is pseudoelastic – sufficiently strong external force is able to restore the orientation from <110>/{111} back to <100>/{001} [19]. The main factor driving the SME is the surface stress induced by the high surface-to-volume ratio of the nanowire [14].

The thickness of the Cu nanowire has a strong influence on the critical temperature of spontaneous reorientation [14], [15], [19] – increasing the thickness from 1.76 nm to 3.39 nm changes the critical temperature from ~100 K to close to the melting point [15]. Although the lattice reorientation process has been directly observed in experiments where Au nanowires were stretched [20], pseudoelasticity and shape memory effect in Cu nanowires remain experimentally unverified [21].

Despite the broad research on metal nanowires, the impact of high electric fields on their stability remains unclear. In [22], the atomistic simulations showed the increase of Young's modulus in <100>, <110> and <111> Ag nanowires in electric field, which may cause the increase in their stability. In [23] it was reported that Cu nanowires melt at decreased temperatures as compared with macroscopic values, which makes it easier for field emission currents to melt such structures, contributing to the decrease in their stability in a strong electric field.

The aim of the current study is to investigate the impact of high electric fields on the stability of Cu nanotips by means of hybrid electrodynamics – molecular dynamics (ED-MD) simulations. This knowledge will contribute to predictive models of surface evolution in such extreme conditions. SME is used as a mechanism for destabilizing the tips, since due to this effect a Cu nanotip of <100>/{001} orientation is capable of shrinking significantly above a critical temperature in a short time. Thus, change in the critical temperature alters the thermal region where <100> wires remain stable, providing in that way an appropriate measure for the change in the stability of a nanotip.

## 2. Methodology

### 2.1. Simulation setup

The molecular dynamics simulations were carried out by means of the open source classical MD code LAMMPS [24] and the Hybrid ED-MD code HELMOD [25], [26], based on the classical MD code PARCAS [27]–[29]. In MD simulations the time step was 4.05 fs and the Mishin [30] embedded-atom (EAM) interatomic potential was used. The potential is constructed in effective pair format and incorporates both experimental and *Ab-Initio* data for fitting. For the purposes of current study it is especially important that the fitting database included the elastic constants of Cu. The potential has been successfully used in previous works [7], [16], [31], has shown good results in lattice defect simulations and accurate reproduction of non-equilibrium system energetics [30].

To investigate the stabilizing and destabilizing effects of an electric field separately, the simulations were carried out in two subsequent stages. At first the entire system – the substrate and a tip on it – were held at the same temperature with Berendsen thermostat [32]. In the second half, the Berendsen temperature control was applied only to the lateral sides of the substrate, and the Joule heating caused by field emission current was added to the nanotip. The additional thermal effects were included to the simulations by scaling the atomic velocities within the tip until the desired temperature distribution was reached. The distribution was found by solving numerically the macroscopic heat equation, taking into account the finite size effects [26].

To ensure correct lattice constants both in bulk and nanosized material without triggering premature shape transformation of the nanotip, the relaxation of the system was done in two consecutive stages. At first, the system was relaxed with energy minimization technique by running the Polak-Ribière [33] version of the conjugate gradient algorithm for 100 successive iterations. After this, the system was instantaneously heated from 0 K to the desired temperature and relaxed at a constant temperature $T = T_{cr}^{est} - 250$ K for 10 ps. The estimated critical temperature $T_{cr}^{est}$ was predetermined by running test simulations, where the temperature was slowly ramped from zero to 1200 K. After the relaxation, the temperature was linearly ramped up with the rate 1.0 K/ps for 400 ps. The ramping rate was chosen as a compromise between computational efficiency and comparability to the previous results [15].

The results of the simulations were visualized with open source program OVITO [34].

### 2.2. Geometry of the simulation cell

In current study a single field emitter is investigated, although in experiments the growth of multiple nanotips is also possible. As shown in [35], the proximity of neighboring nanotips results in effective screening of local electric field. The screening phenomenon together with surface charge induced on top of nanotips may affect the processes studied in the present work. Such interactions, however, were considered negligible within the chosen timespan and distances.

Figure 1 illustrates the geometry of the simulated system. A single-crystalline cylindrical nanotip with a hemispherical cap and smoothened cylinder-substrate junction was placed on the monocrystalline Cu substrate. Similar geometry was observed in the Au nanopillar growth experiment in [36]. Initial dimensions of the tips were 6, 8, and 14 nm in height and the diameters varied in the range from 1.6 to 2.8 nm. The dimensions of substrate were chosen to be 23×23×1.4 nm³. The height and diameter of the tip were selected to match the previous studies [5], [13], [37] and to result in critical temperatures below the melting point of the tip. The system was periodic in the lateral directions, its top surface was exposed to vacuum and two bottom atomic layers of the simulation cell were fixed in place. Both the tip and the substrate were cut out from the single-crystalline Cu block with the lattice constant of 3.61 angstroms. The block was initially <100> oriented in axial and {001} oriented in lateral directions.

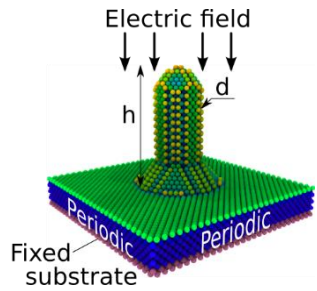

Figure 1. Geometry of the simulated system. On the substrate (blue atoms) with dimensions 23×23×1.4 nm³ there is a cylindrical nanotip with height $h$ and diameter $d$. Two bottom atomic layers of substrate, that are colored red, are fixed. The system is periodic in lateral directions. The top yellowish-green surface is exposed to vacuum and electric field. The geometric proportions on the image differ from the ones used in simulations.

### 2.3. *Electric field*

To obtain the distribution of electric field above the tip, the Laplace equation with mixed Dirichlet and Neumann boundary conditions was solved. As boundary conditions, uniform electric field was applied to the top of the simulation cell, while constant potential was used to model the copper surface. Due to the external field the partial charges are induced on the surface atoms that in turn make them interact with the electric field. Charges on surface atoms were calculated from the values of the local field near the atoms, applying the Gauss's law as given by the "pillbox" technique [8]. The resulting Lorentz and Coulomb electric forces were incorporated into Newtonian atomic motion as described in [25].

Due to geometric effects, the local electric field near the nanotip is stronger than the applied field. The field enhancement factor $\beta$ of the tip equals $\beta \approx 2 + h/r$ [38], where $h$ is the height of the emitter and $r$ is its radius. In the limit of $h \gg r$, the overall field enhancement factor is determined by $h/r$ ratio.

Due to field enhancement, the applied electric fields used in the current work become strong enough to initiate a field emission current from the metal tip [39] that may cause significant increase in tip temperature via resistive heating [10]. To calculate the field emission current, the tip was divided into separate vertical pillars of identical cross-sections and varying heights. The current in each pillar was calculated using general thermal-field emission equation, taking into account the local electric field strength above each pillar and the temperature of the field emitter [40], [41]. The total current value was obtained by summing up all the discrete currents [26].

Experimentally it is well established that electric fields of ~10-11 GV/m can cause field evaporation of Cu surface atoms [5]. In our simulations the *h/r* of a tip and the strength of external electric field did not exceed 10 and 1.5 GV/m, respectively, which ensured that the values of local electric fields in simulations were safely below the critical fields.

### *2.4. Critical temperature of spontaneous reorientation*

In [14]–[16], the critical temperature above which a <100>/{001} nanowire reorients spontaneously into <110>/{111} configuration was determined by considering the average atomic potential energy of the wire. The wire with <100>/{001} orientation has higher energy than a <110>/{111} one and the initiation of the reorientation process is marked by a strong decrease in the potential energy. This approach is clearly applicable if the length of the wire is at least few tens of nanometers, to avoid the influence of statistical fluctuations and dependence on initial conditions. The current work focuses on nanotips that may grow on metal surface under high electric field. The number of atoms in such systems is much smaller and the drop of potential energy due to reorientation is less significant than in previous works. Such an issue can be surpassed with a common neighbor analysis (CNA) based technique that was validated with the cases where both potential energy and CNA-method were usable.

### *2.5. Common neighbor analysis technique*

In the current work, the common neighbor analysis (CNA) method is proposed to determine the critical temperature of spontaneous reorientation. The method is based on the observation that the reorientation of the tip is always preceded by nucleation of a dislocation. The atoms forming a stacking fault between the two Shockley partial dislocations are detected with the CNA algorithm [42] as atoms of hexagonal close-packed (HCP) structure. The CNA algorithm detects the local crystal structure around every atom by analyzing the bond length between a given atom and its neighbors. The cutoff radius 3.08 angstroms was chosen for this tool as a midpoint between the first and the second shells of neighbors in FCC crystal. Thus, the critical temperature of spontaneous reorientation is in the current work defined as a temperature where the amount of atoms in a system forming the dislocation (dislocation signal) increases sharply.

To reduce the noise in the dislocation signal, two-step filtering was applied to it. On the first stage, small fluctuations in the beginning of the signal were suppressed to a constant value. After that, band pass filter was applied to the Fourier spectrum of the signal as proposed in [43]. Denoting the dislocation signal as *f*, its Fourier transform as *F* and the inverse Fourier transform as $F^{-1}$, the signal after filtering looked like

$$f_{out} = F^{-1}(F_{in} \cdot G), \tag{1}$$

where the filter function $G$ with filter parameters $\tau_1$ = -7.0 and $\tau_2$ = -60.0 was chosen to be

$$G_i = \exp\left(\tau_1 \cdot \frac{i}{N}\right) + \exp\left(\tau_2 \cdot \frac{N-i}{N}\right), \qquad i = 0, 1, 2, \ldots N. \tag{2}$$

The filter function was chosen as a rough compromise between the ability to reduce the noise by discriminating the middle frequencies, and the ability to follow the sharp changes in dislocation signal by passing the low and high frequencies. The sharp changes in the filtered signal were detected by its time derivative. To reduce the noise, derivative values lower than half of the maximum were suppressed to zero. In that way the reorientation signal showing the changes in reorientation process is obtained.

The reorientation often has several phases that can be seen as the stages in dislocation signal. For that reason there is sometimes more than one peak in a reorientation signal. Experimenting with different tips and initial conditions showed that usually no more than three first peaks correspond to the changes in reorientation process. The algorithm is able to distinguish the following main stages of the reorientation process: the formation of dislocations, the start of reorientation, and the change in reorientation rate. For consistency, the critical temperature in different simulation geometries and initial conditions was determined by the first peak in the reorientation signal.

### 3. Results and discussion

#### 3.1. *Determining the critical temperature*

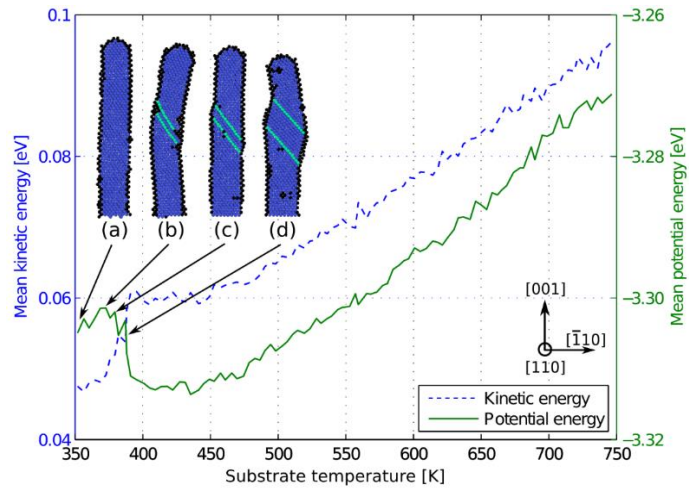

Figure 2. The change of mean atomic kinetic and potential energy of the system. The solid green and dashed blue lines depict mean potential and kinetic energy of atoms in a tip, respectively. The evolution of partial dislocations in a tip is illustrated with the snapshots (a) to (d) which are colored according to the local crystal structure (dark blue – FCC, light blue – HCP, black – other) as determined by common neighbor analysis, and oriented as shown by the axes above the legend.

Figure 2 depicts the change in mean atomic potential (solid green line) and kinetic (dashed blue line) energy in a nanotip on a heated substrate. The snapshots of the cross-sections of the tip at the substrate temperatures 350-750 K, colored according to the local crystal structure as determined by common neighbor analysis [42], are placed as insets (a) to (d).

At temperatures below critical, both potential and kinetic energy inside the tip grow gradually with increasing temperature (line segment between snapshots (a) and (b)). Once the temperature reaches a certain value, a $\frac{1}{6}\langle 112 \rangle$ Shockley partial dislocation nucleates at the side surface of the tip [31]. This propagates through the tip on a {111} plane, until the stacking fault caused by the partial dislocation extends through the entire cross-section of the tip. The dislocation reaction is completed by the nucleation of the Shockley partial dislocation reflection (figure 2(b)), after which they start to glide in opposite directions through the tip on a {111} plane (figure 2(c)). The partial dislocation and its reflection leave behind the twin boundaries that separate initial and reoriented parts of a tip. Through the glide of twins the tip reorients gradually from <100>/{001} into <110>/{111} configuration that results in a decrease in potential and increase in kinetic energy (figure 2(d)).

We noticed that on average, there is a thermal separation between the nucleation of the first partial dislocation and the beginning of reorientation is, which amounts to 20-60 K. The temporal separation between those events is in the order of picoseconds. The research also showed that during the reorientation process the amount of partial dislocations changed and, in some cases, a nanotip was found tilted after reorientation (figure 3).

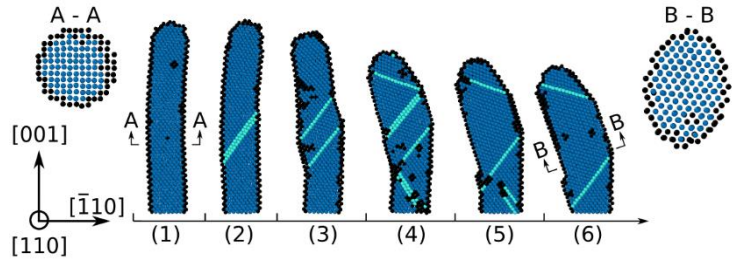

Figure 3. The partial dislocation in the $h = 14$ nm, $d = 2.4$ nm nanotip at instances indicated by the bracketed numbers in figure 4(a). Cross-sections A-A and B-B demonstrate the atomistic packing of initial and reoriented nanotips, respectively. Tips are colored according to the local crystal structure (dark blue – FCC, light blue – HCP, black – other) as determined by common neighbor analysis, and oriented as shown by the axes on the left.

The extent of reorientation varies strongly with different initial conditions. If the tip reorients in its whole length, the drop in potential energy becomes clearly visible, as shown on figure 4(a). However, in case of small extent of reorientation, the drop in potential energy becomes almost invisible and the changes in the system are not revealed, as demonstrated on figure 4(b). On the other hand, the partial dislocations appear every time the temperature is high enough and does not depend on initial velocities of atoms. After smoothing the signal of the amount of atoms forming the partial dislocations (dashed red lines on figure 4), its truncated derivative (long-dashed magenta lines) could be used to pinpoint the critical temperature.

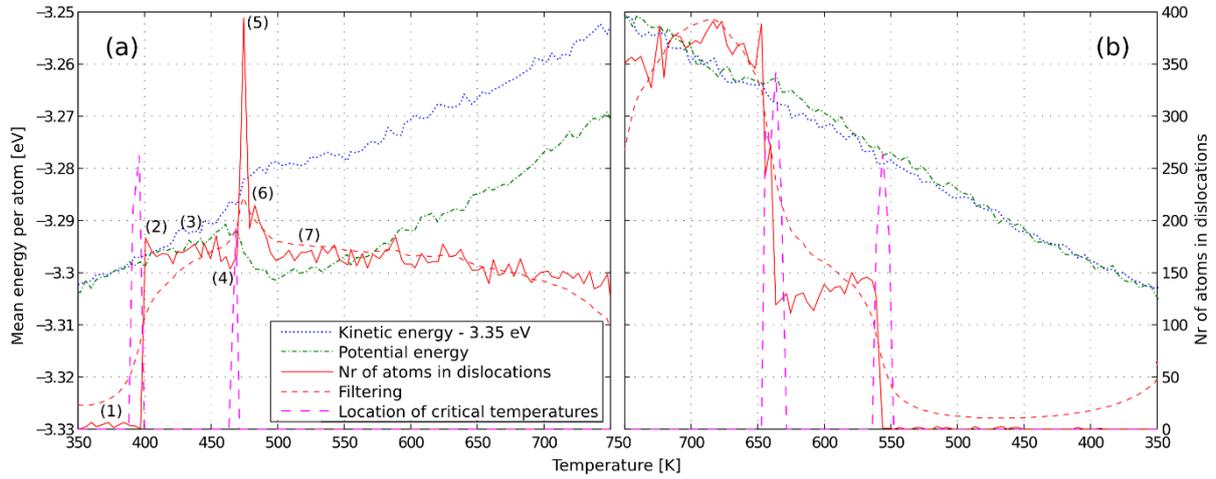

Figure 4. The mean atomic potential energy, kinetic energy and number of atoms in dislocation as a function of temperature. Dotted blue and dash-dotted green lines represent mean atomic kinetic and potential energy of a tip, solid red line shows the number of atoms in dislocations and dashed red is the result of smoothing the latter via filtering. The long-dashed magenta line expresses the upper half of the derivative of smoothed data and is used to pinpoint the location of critical temperatures. Kinetic energy is shifted down by 3.35 eV to fit to the graph. (a) and (b) represent the same system with different initial atomic velocities. Points (1) to (6) mark the instances where snapshots on figure 3 were extracted.

*3.2. Validation of critical temperature measurements*

The critical temperature measurements were validated by comparing the data obtained in the current study with the previous results by Liang et al [15]. Figure 5 shows how the critical temperature of spontaneous reorientation depends on the tip diameter with various temperature ramping rates if no electric field is present. It can be seen that exponential decrease in ramping rate results in roughly linear drop of the $T_{cr}$ for each given size of the tip. The critical temperatures on figure 5 gradually approach the values reported in [15], where the system was allowed to evolve at a given constant temperature for a sufficiently long time. The ramping rate in the current work was chosen to be slow enough to describe sufficiently well the reorientation process, and appropriately high to allow reasonable computational efficiency in the statistical analysis of the studied phenomenon.

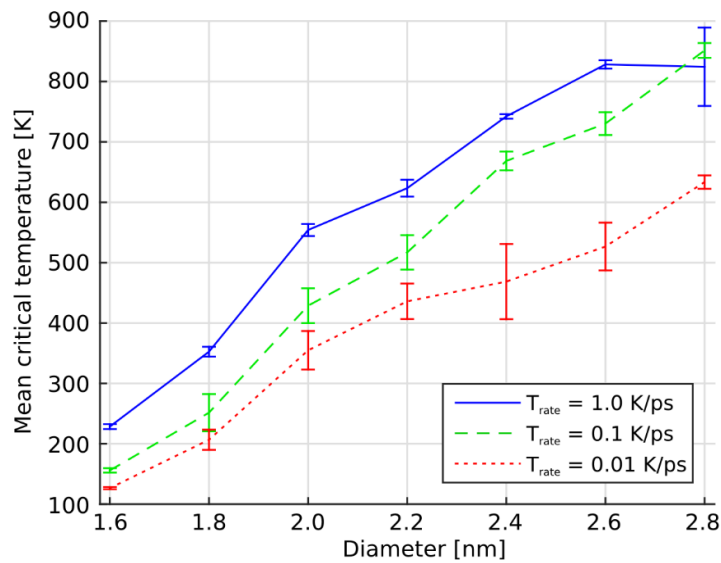

Figure 5. The change of mean critical temperature with different temperature ramping rates. Error bars show the standard mean error for six independent simulations.

### 3.3. The change of critical temperature due to applied electric field

Figure 6 shows the change in critical temperature of spontaneous reorientation in the 6 nm Cu nanotip with different diameters in various electric fields. It reveals that the application of a strong electric field increases the critical temperature by hundreds of kelvins, increasing the stability of <100>/{001} tips. It is, however, important to mention that the electric field without field emission current effect behaves as a stabilizer only up to the critical electric field values, which can cause field evaporation of atoms. In copper, these values range between 10 and 14 GV/m [5], [10].

As shown on figure 6, $T_{cr}$ is proportional to the tip diameter and thus inversely proportional to its surface-to-volume ratio. It means that in <100> tips with lower surface-to-volume ratio, higher temperature is needed for a dislocation to be nucleated. This is in line with conclusions in [14], [15] where the reorientation was found to depend on the surface stress.

It is worth noting that the critical temperatures become less electric-field-dependent as the diameter of tip increases. This is because the strength of the local electric field near the tip apex, which causes the stabilizing effect on the tip, depends on the geometric aspect ratio as $E_{loc} \approx h/r \cdot E$. Since the increase in the diameter of the tip of a given height leads to the decrease in its aspect ratio, the local enhanced value of the applied electric field near the tip apex is also lowered.

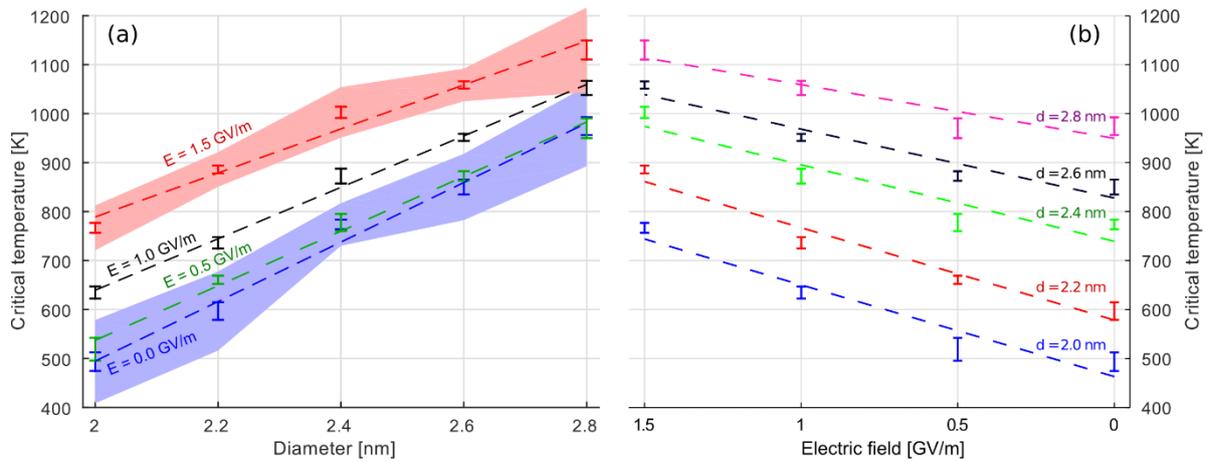

Figure 6. Mean critical temperature dependence on tip diameter in various electric fields without Joule heating from field emission current. Error bars represent standard mean error for 20 simulations with different initial atomic velocities; dashed lines are their linear fit; colored regions show the variation of critical temperature within one standard deviation. For the sake of clarity the variation within one standard deviation is shown only for minimum and maximum fields. Critical temperature variation is shown both in diameter domain (a) and electric field domain (b).

### 3.4. Effects of field emission current on critical temperature

The effect of electron emission current in terms of Joule heating on the value of critical temperature of spontaneous reorientation was investigated in the 8 nm nanotips. Figure 7(a) shows the decrease in $T_{cr}$ due to the impact of field emission current as compared to the case without it. As can be seen, the higher the applied field, the more the critical temperature is lowered. The field emission current is an exponential function of the local electric field and ranges between 0.2-7 pA ($E$ = 0.4 GV/m) and 0.6-3 μA ($E = 1.0$ GV/m) in the tips with diameters 2.4-1.6 nm. The corresponding current densities are 0.3-20 pA/nm$^2$ and 0.9-7 μA/nm$^2$, respectively. Even the highest current density lowers the critical temperature of spontaneous reorientation by a few tens of kelvins, at most by 50 K.

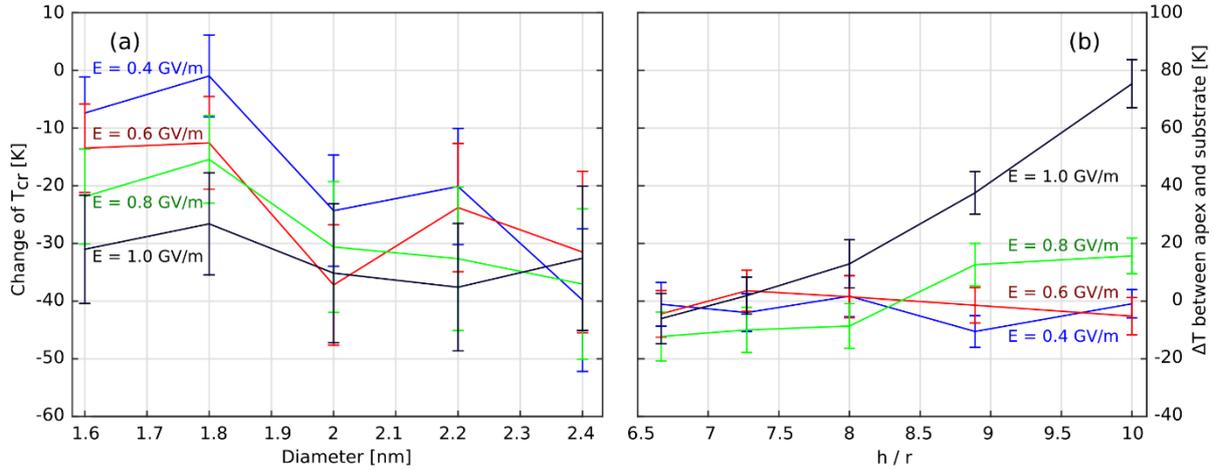

Figure 7. (a) The difference between critical temperatures with and without the effects of field emission current as a function of field emitter diameter. (b) Difference between apex and bottom temperatures of a tip at the start of reorientation as a function of aspect ratio. Error bars are the standard mean error for 105 independent simulations.

The drop in the critical temperature can be explained by the non-uniform temperature distribution in the tip. At the very top, where the cross-sectional area is the smallest, the current density reaches the highest values and due to finite heat conductivity, the tip apex becomes significantly hotter than the bottom. Because of non-zero temperature gradient, the dislocations typically form within relatively narrow apex area when the local temperature reaches sufficiently high value to allow for a reorientation event to occur. In our simulations the critical temperature was estimated as an average temperature within the entire tip at the moment the first dislocations nucleated. Averaging the hot apex and the warm bottom resulted in cooler values for the overall $T_{cr}$. A typical temperature distribution along the 8 nm tip in the beginning of reorientation is shown on figure 8.

As can be seen on figure 7(b), the difference between the temperatures in the tip apex and the bottom is larger in the nanotips with higher aspect ratios $h/r$. The difference also strongly depends on the electric field strength – in a tip with $h/r = 10$, the temperature difference is ~0 K, 16 K and 75 K for applied electric fields of 0.6 GV/m, 0.8 GV/m and 1.0 GV/m, respectively.

Hence it can be expected that increasing the tip aspect ratio over some limit may compensate for the stabilizing effect of the field itself. This, however, was not observed in our simulations, because the apex temperature reached its melting point before such destabilization occurred.

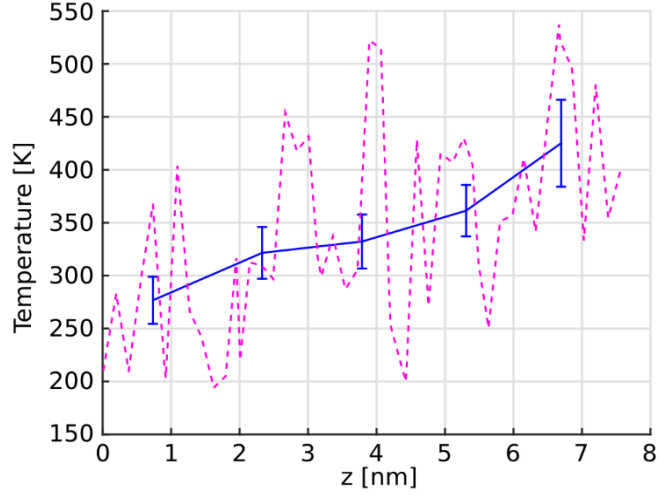

Figure 8. Vertical temperature distribution in the 8 nm nanotip under applied electric field of 1.0 GV/m. Dashed magenta line shows the variation of temperature, solid blue depicts the temperature average and its standard mean error for 1/5-th of a tip.

### 3.5. Stability of nanotip and local field enhancement factor

As a result of the reorientation, a 6 nm <100> tip may lose up to 20-27% of its initial height. Although it is about half as much as reported for long nanowires [16], such a significant loss in the major dimension will strongly affect the local field enhancement factor $\beta \approx h/r$ [38] of the applied electric field, where $h$ is the height of a tip and $r$ its radius. Assuming that the volume of a tip remains constant during the reorientation process, it can be shown via basic calculus that some loss in relative height causes equal gain in relative cross-sectional area. Considering the relative change in field enhancement factor $d\beta/\beta$ to be equal to the difference between relative change in height and radius, $d\beta/\beta = dh/h - dr/r$, the relative loss in field enhancement factor for a tip with circular cross-section turns out to be $\frac{d\beta}{\beta} = 1.5 \cdot \frac{dh}{h} = 1.5 \cdot 23\% \approx 34\%$.

The electric field applied to the tip increases the $T_{cr}$ and hence broadens the temperature range where <100>/{001} nanotips remain stable and retain their full length for a relatively long period of time. Moreover, in [15] it was shown that a tensile stress of ~5 GPa applied to the 1.76×1.76 nm$^2$ Cu <110>/{111} wire at 200 K induces its reorientation into <100>/{001} configuration. A tip can experience such tensile stress in the presence of a rather high local electric field (~20 GV/m [8]). If such reorientation takes place through the whole length of the nanotip, it may pseudoelastically stretch out up to 50% [16] and gain in constant volume approximation about 75% in aspect ratio.

Thus, based on the previous considerations, estimated fluctuations in $\beta$ can be up to -34% to +75% from its average value. This is in a good agreement with the experimental results. In [5] it was found that the breakdown electric field and the field enhancement factor fluctuated roughly 35% around their average values. Similar conclusions can be drawn from conditioning curves obtained for Cu, stainless steel, W, Cr, Ti, Nb and other metals [37].

Such a comparison is not sufficient to claim that pseudoelasticity and shape memory effect alone are the responsible mechanisms for the appearance and vanishing of high aspect ratio field emitters on Cu surfaces under high electric field. It provides, however, another possible explanation to the inconsistency between the experimental measurements of the field enhancement factor [5] and microscopy results [11]. For example, after the <100>/{001} to <110>/{111} reorientation, the flattening of the remaining emitter may take place in a matter of seconds, as described in [12], enhancing the overall $\beta$ fluctuation range. Moreover, the field emitter may be located on top of already existing surface roughness. In that case the field enhancement factors will be multiplied due to Schottky's conjecture amplifying the fluctuation of $\beta$ even more [44].

## 4. Conclusions

The current work investigated the impact of high electric fields on the stability of <100>/{001} Cu nanotips that may grow spontaneously on Cu surfaces in the presence of electric fields. The shape memory effect was utilized to destabilize the Cu nanotips. Due to SME the Cu tip may spontaneously reorient from <100>/{001} into <110>/{111} orientation, along with significant decrease in its height. The reorientation takes place only at temperatures higher than the critical one. The change in critical temperature of spontaneous reorientation on high electric field was used as a measure of the change in <100>/{001} tip stability.

It was shown that strong electric field has both stabilizing and destabilizing effect on Cu nanowires. Local electric field of less than 10 GV/m induces tensile stress to the system that is weak enough not to break, but sufficient to stabilize the system. The resistive heating caused by the field emission current is the main destabilizing factor. As a result of the study it was found that applied electric fields above 1 GV/m increased the critical temperature of $h$ = 6-8 nm, $d$ = 2.0-2.8 nm <100> Cu tips by hundreds of kelvins and made them more stable. The effects of emission current lowered the $T_{cr}$ in those field emitters by tens of kelvins due to uneven temperature distribution in the tip. In applied electric fields higher than 1 GV/m, the tips with $d \gg 2.8$ nm or $h \gg 8$ nm showed apex temperatures that exceeded their melting points and were left out of the scope of the current paper.

The shape memory effect is one plausible mechanism that could explain the lack of experimental evidence for Cu nanotips with a diameter less than 2.8 nm. Due to SME, a <100>/{001} tip that is stable under strong electric field, becomes unstable without it and may contract up to 23% of its initial length. Assuming the constant volume in high and thin tip, the spontaneous reorientation was shown to cause a drop up to 34% in the field enhancement factor of the system. Such variation agrees well with the previous experimental work.

During the study a novel algorithm was developed to pinpoint the critical temperature of spontaneous reorientation in small nanotips. The algorithm uses common neighbor analysis to spot the formation of dislocations in the system that mark the start of the reorientation process. The method is compatible with the techniques that consider the change of potential energy, and is usable in a wide range of nanowires. The algorithm also enables to pinpoint the different stages in the reorientation process.

**Acknowledgments**

The current study was supported by the ''Research internationalization'' program of the European Regional Development Fund, Academy of Finland project AMELIS, Estonian Research Council Grant PUT 57 and the national scholarship program Kristjan Jaak, which is funded and managed by Archimedes Foundation in collaboration with the Ministry of Education and Research of Estonia. The authors thank the IT Center for Science Ltd. (CSC) and the Finnish Grid and Cloud Infrastructure for providing the computation time.